# SKIPNet: Spatial Attention Skip Connections for Enhanced Brain Tumor Classification

Khush Mendiratta[1][0009-0001-8768-4155], Shweta Singh[2], Pratik Chattopadhyay[2][0000-0002-5805-6563]

[1] Indian Institute of Technology Roorkee, India
[2] Indian Institute of Technology BHU, India
khush_m@ma.iitr.ac.in, shwetasingh.rs.cse20@itbhu.ac.in, pratik.cse@iitbhu.ac.in

**Abstract.** Early detection of brain tumors through magnetic resonance imaging (MRI) is essential for timely treatment, yet access to diagnostic facilities remains limited in remote areas. Gliomas, the most common primary brain tumors, arise from the carcinogenesis of glial cells in the brain and spinal cord, with glioblastoma patients having a median survival time of less than 14 months. MRI serves as a non-invasive and effective method for tumor detection, but manual segmentation of brain MRI scans has traditionally been a labor-intensive task for neuroradiologists. Recent advancements in computer-aided design (CAD), machine learning (ML), and deep learning (DL) offer promising solutions for automating this process. This study proposes an automated deep learning model for brain tumor detection and classification using MRI data. The model, incorporating spatial attention, achieved 96.90% accuracy, enhancing the aggregation of contextual information for better pattern recognition. Experimental results demonstrate that the proposed approach outperforms baseline models, highlighting its robustness and potential for advancing automated MRI-based brain tumor analysis.

## 1.Introduction

In 2016, brain tumors emerged as the leading cause of cancer-related deaths among children aged 0–14 in the United States, surpassing leukemia [1]. Additionally, brain and central nervous system (CNS) tumors ranked as the third most common cancer in teenagers and young adults aged 15–39 [2]. Different types of brain tumors necessitate various medical interventions, highlighting the need for accurate diagnostic systems.

Conventional computer-aided diagnosis systems require the identification and segmentation of the tumor mass before it can be classified into different types. After segmenting the tumor, the next steps involve feature extraction and classification. Recent studies on brain tumor identification and segmentation [3-4]have indicated that there is no universal system capable of accurate tumor detection, irrespective of location, shape, or intensity [5].

Numerous algorithms have been proposed for feature extraction and classification of brain tumors. The grey-level co-occurrence matrix (GLCM) is widely used for extracting low-level features [6–8]. Other feature extraction techniques that address the complex texture of brain tumors include neural networks [8, 9], the Bag-of-Words (BoW) model [10, 11], and Fisher vectors [12]. A notable study demonstrated that combining adaptive spatial pooling with the Fisher vector algorithm allows for the classification of brain tumors into categories such as glioma, meningioma, and pituitary tumors, achieving accuracies ranging from 71.39% to 94.68% [12].

An additional issue with the presented CNN models is that they require an enormous number of learnable parameters. In this work, we propose a novel attention-based SKIPNet for brain tumor classification that addresses these issues.

The following is an outline of this paper's direct contributions:
1. Enhanced Feature Extraction with Spatial Attention.
2. Improved Spatial Context Integration.
3. Consistent Application of Spatial Attention.
4. Efficient Feature Map Reduction and Classification.

The rest of the paper is structured as follows: Section 2 describe the related work. Section 3 describes the proposed system's methodology, including a complete phase description. Section 4 contains the results with analysis and the comparison research. Lastly, Section 5 includes closing remarks.

## 2.Literature Review

The study of brain tumor classification using various advanced techniques has yielded impressive results, as highlighted by several authors in recent years. Irmak et al.[13] utilized a fully optimized convolutional neural network, achieving an accuracy of 94.7%, although their model is limited to brain tumors. Liu et al. [14] focused on meningioma classification with a remarkable accuracy of 97.5% through a multi-contrast MRI approach, but faced challenges due to the limited dataset size. Mao et al. [15] employed a hybrid method combining convolutional neural networks and gradient boosting decision trees, achieving 94.7% accuracy, though the complexity of their model architecture is a downside.

Qian et al. [16] also explored a hybrid approach using wavelet transforms with CNNs, reaching an accuracy of 94.1% but sharing similar architectural complexities. Jingwei et al. [17] achieved 94.8% accuracy in tumor segmentation and classification with a U-Net architecture, yet their work is confined to a small dataset of brain tumors. Talo et al. [18] implemented a transfer learning method based on Inception-ResNet-v2, achieving 94.0% accuracy despite limited interpretability.

Alom et al. [19] utilized a hybrid CNN-RNN approach for tumor detection, achieving an accuracy of 96.3%, but their model is primarily suited for classification and detection, lacking segmentation capabilities. Nandy et al. [20] achieved an impressive accuracy of 98.27% through a transfer learning approach, though their sample size was limited. Gurleen [21] and Sahoo [22] proposed hybrid approaches combining PSO with support vector machines (SVM) and multi-layer perceptrons (MLP), respectively, achieving accuracies of 96.6% and 91.3% but with limited comparative analyses. Overall, while these studies present significant advancements in brain tumor classification, challenges such as limited datasets and model complexity remain critical issues to address. The proposed model outperformed the others in a number of parameters, reaching a maximum test accuracy of 96.90% with a 0.1435 loss. Better contextual information aggregation was achieved by integrating spatial attention with the conventional Conv-Nets design. This was essential for identifying complex patterns in the MRI images.

## 3.METHODOLOGY

The proposed model architecture, termed SKIPNet, integrates CNN Blocks and a Spatial Attention Layer (SAL) to enhance the feature extraction process for brain MRI images. Each CNN Block comprises two main sub-components: the Spatial Attention and Sub-block. These blocks utilize convolutional layers, batch normalization, and spatial attention mechanisms to progressively refine and capture crucial spatial and contextual information. The SAL is designed to assign importance to different spatial regions of the feature map, helping the network focus on abnormalities or specific anatomical features that vary significantly across MR images. The Spatial Attention Layer (SAL) is a mechanism designed to enhance the model's ability to focus on important regions of the feature map by assigning varying levels of attention to different spatial areas. This allows the network to prioritize areas that are more likely to contain abnormalities or significant anatomical features, which can vary across different MRI images. By emphasizing these critical regions, the SAL helps the model improve its accuracy in detecting and classifying brain tumors, as it can more effectively capture the spatial variations and subtle patterns that are crucial for accurate diagnosis.

This strategy is particularly beneficial for tasks like brain tumor detection, where spatial variability poses significant challenges. The Spatial Attention Layer (SAL) emphasizes critical spatial locations by learning an attention map. Given an input feature map ,SAL reduces its channel dimensions using a 1×1 convolution, followed by 3×3 convolutions with dilation, and final reduction to a single channel via another 1×1 convolution. The resulting attention map is normalized and element-wise multiplied with the original input feature map to enhance the representation of significant regions. This layer's ability to dynamically adjust feature weighting helps in consistently extracting relevant spatial features critical for detecting lesions and abnormalities in MR images. The CNN Block combines the strengths of spatial attention and traditional networks. The Spatial Attention Sub-block refines feature maps by emphasizing important spatial regions, while the Convolutional Sub-block performs deep feature extraction using a series of convolution and batch normalization layers. The two sub-blocks work synergistically, with the attention sub-block refining the focus and the convolutional sub-block extracting rich hierarchical features. The output of these sub-blocks is integrated and subjected to max-pooling, yielding a robust feature map that balances spatial refinement and deep feature extraction. The SKIPNet architecture employs a sequence of four CNN Blocks, each followed by a dropout layer to prevent overfitting. This iterative approach enables the model to learn increasingly complex features for identifying abnormalities in MRI scans. The SKIPNet architecture is designed to learn progressively complex features for identifying abnormalities in MRI scans through a sequence of four CNN Blocks. Each of these blocks is responsible for extracting different levels of spatial information from the input image, enabling the model to focus on both fine and coarse details. By applying multiple convolutional layers within each block, the architecture is able to capture increasingly intricate patterns and structures, which are critical for detecting variations in brain anatomy, such as tumors. This hierarchical approach ensures that the network can learn a robust representation of the MRI scans, improving its ability to discern subtle abnormalities.To prevent overfitting, a dropout layer is applied after each CNN Block. The dropout layer works by randomly disabling a fraction of the neurons during training, forcing the network to rely on a wider range of features and preventing it from becoming overly specialized to the training data. This regularization technique enhances the model's generalization ability, ensuring that it performs well on unseen data and is less likely to memorize irrelevant details. Together, the combination of the CNN Blocks and dropout layers enables SKIPNet to effectively learn complex patterns while maintaining high accuracy and robustness in identifying abnormalities in MRI scans.

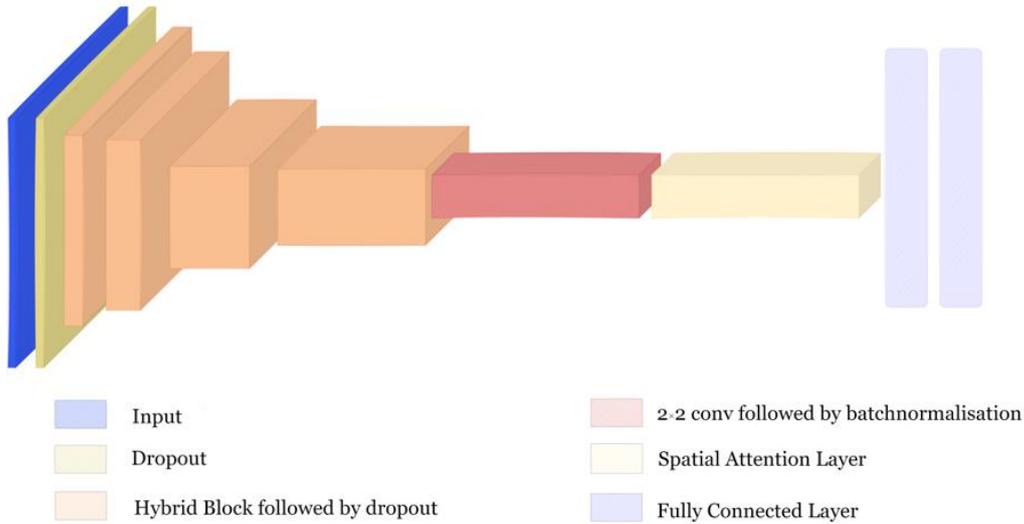

A 2×2 convolutional layer is used to further downsample the feature map, reducing computational complexity while retaining critical information. Before the fully connected layers, a SAL is applied again to maintain spatial awareness during classification. This consistent use of SAL ensures that the model effectively captures and utilizes spatial information throughout the network, from early feature extraction to the final classification stage. By applying spatial attention at multiple stages and integrating dropout for regularization, SKIPNet demonstrates enhanced robustness and generalization. This design ensures a coherent approach to feature representation, leveraging both low-level and high-level spatial features for accurate brain MRI classification. The combined emphasis on spatial refinement, deep feature extraction, and spatial awareness in fully connected layers enables SKIPNet to effectively classify MR images, achieving a balanced trade-off between complexity and accuracy.

## 4.Experimental Result and Analysis
### 4.1 Dataset used
The algorithm SKIPNet is trained on a dataset obtained through Figshare [23]. The brains of people with brain tumours were scanned using MRI in the creation of this dataset. The dataset's salient details are as follows:
**Data Source**: Figshare [24], a website for exchanging research data and scientific outputs, is used to access the dataset.
**Image Metadata**: Each MRI image in the collection is accompanied by a text file including patient-specific metadata, such as the patient's gender, age, and tumour type.
**Image Format**: The dataset's images are kept in the widely used DICOM (Digital Imaging and Communications in Medicine) format, which is used for medical imaging. Applications in healthcare and medical research can benefit from the standard medical image storage and transmission format, DICOM.

**Table 1. Dataset description**

| Tumor Category | Meningiomas | Gliomas | Pituitary | Total |
|---|---|---|---|---|
| Number of patients | 82 | 91 | 60 | 233 |
| Number of slices | 708 | 1426 | 930 | 3064 |

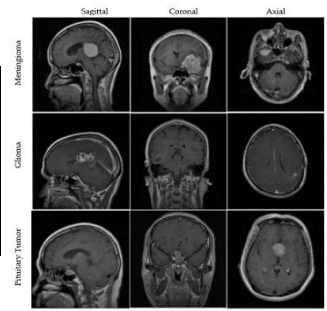

### 4.2 Experimental Setup
This section outlines the configuration details for the proposed method. We implemented deep neural networks optimized through an evolutionary process using Python. The models were developed in a Python 3 environment, utilizing the Keras framework with a TensorFlow backend. To leverage computational power, we used Kaggle's GPU resources, which include the following specifications:

- GPU: P100 with 2496 CUDA cores and 16 GB of high-memory virtual machines.
- CPU: A single hyperthreaded Xeon processor running at 2.3 GHz (1 core, 2 threads), with a 45MB cache.
- RAM: 25 GB.

These specifications provided the necessary resources to effectively train and test the deep learning models.

### 4.3 Metrics
**Loss/Cost function**
Sparse categorical cross-entropy loss has been used to monitor the training process. It is a variation of the cross-entropy loss function that's used for multi-class classification tasks. It's used when the target labels are integers instead of one-hot encoded vectors.

$$loss = -\sum_{i=1}^{n} y_i * \log \hat{y}_i \qquad (7)$$

### Evaluation parameters
<u>Accuracy</u>: - it is the main metric which has been used to monitor the performance of model at every epoch and also in comparison with other models. It is defined as number of instances classified correctly per hundred instances. It is given by:

$$Accuracy = \frac{(TP + TN)}{(TP + FP + FN + TN)} \qquad (8)$$

where TP represents the true positives and TN represents the true negatives, while FP is false positives and FN is false negatives of each class.

### 4.4 Results and Discussion
Table 2 compares our work with state-of-the-art architectures and establishes the superiority of SKIPNet. It outperforms all SOTA methods in both accuracy and loss minimization and has the smallest size among all.

**Performance Overview**
The proposed model exhibits a remarkable accuracy of 96.90%, which significantly surpasses the performance of all benchmark models tested. This accuracy represents a substantial improvement over the next best-performing model, ResNet-50, which achieves an accuracy of 95.07%. The increase of approximately 1.83% in accuracy reflects the proposed model's superior ability to correctly classify instances from the dataset. This level of accuracy is indicative of a highly refined model that can potentially offer better decision-making capabilities in practical applications. The proposed model demonstrates an impressive accuracy of 96.90%, outperforming all benchmark models tested in the study. This exceptional performance marks a significant leap over the next best-performing model, ResNet-50, which achieves an accuracy of 95.07%. The proposed model's accuracy exceeds ResNet-50 by approximately 1.83%, highlighting its enhanced ability to accurately classify instances from the dataset. This improvement indicates that the proposed model has a stronger capacity to detect and differentiate key features in MRI scans, which is crucial for tasks like brain tumor detection. This level of accuracy underscores the model's robustness and potential for real-world applications. The refined architecture and learning approach allow the model to make more precise decisions, making it highly suitable for practical use in clinical settings. With its superior performance, the proposed model could aid in more reliable and accurate diagnoses, supporting healthcare professionals in the early detection of brain tumors and improving overall patient outcomes.

**Loss Analysis**
Loss is a critical metric that quantifies the model's error in predictions. The proposed model achieves an exceptionally low loss of 0.1435, which is significantly better than the losses reported by existing models. For instance, MobileNet, which has the second-lowest loss at 1.1297, still demonstrates a higher error rate compared to the proposed model. The drastic reduction in loss not only indicates that the proposed model is more accurate but also highlights its enhanced ability to generalize well to new, unseen data. This lower loss implies fewer misclassifications and more reliable predictions, which is crucial for applications requiring high precision. In proposed model's superior performance can be attributed to the integration of spatial attention mechanisms with the traditional Conv-Nets architecture, which enhances its ability to capture and emphasize relevant spatial features in brain MRI images. This innovative approach leads to more accurate classifications and lower prediction errors compared to traditional models. Loss is an essential metric in machine learning that measures the error in a model's predictions, indicating how closely the model's output aligns with the true values. In the context of brain tumor detection, a low loss is crucial for ensuring the model makes accurate predictions and avoids misclassifications. The proposed model achieves an exceptionally low loss value of 0.1435, significantly outperforming existing models. For example, MobileNet, which has the second-lowest loss of 1.1297, exhibits a considerably higher error rate than the proposed model. The drastic reduction in loss showcases the model's superior ability to minimize prediction errors, reflecting its greater accuracy and precision. This improvement is vital in applications where the model's ability to provide correct diagnoses is paramount, particularly in medical imaging tasks such as MRI analysis.The notably low loss also highlights the proposed model's ability to generalize well to new, unseen data, a key factor in evaluating the effectiveness of machine learning models in real-world scenarios. A model that overfits the training data might perform well on known datasets but fail to make accurate predictions on unfamiliar data. The proposed model, with its low loss, demonstrates that it not only fits the training data effectively but also maintains high performance when exposed to new examples, making it reliable for real-world use. This generalization capability is critical in the medical field, where the model must be capable of accurately analyzing a wide variety of patient MRI scans, which may differ due to factors such as patient demographics, tumor types, and image quality. The superior performance of the proposed model can largely be attributed to its integration of spatial attention mechanisms with a traditional convolutional neural network (Conv-Nets) architecture. The spatial attention layer allows the model to focus on the most relevant regions of the MRI images, emphasizing spatial features that are critical for tumor detection. This ability to direct attention to specific anatomical features and abnormalities improves the model's classification accuracy. By enhancing the model's capacity to capture and prioritize important spatial information, the integration of spatial attention helps reduce prediction errors and misclassifications, setting the proposed model apart from traditional Conv-Nets that might struggle to distinguish subtle patterns in complex MRI scans. This innovative approach leads to more accurate and reliable predictions, demonstrating the model's potential for practical applications in medical diagnostics.

**TABLE 6.** Comparative analysis with CNN-based SOTA BTC approaches on Figshare Brain MRI dataset.

| Reference | Accuracy | Approach |
|---|---|---|
| Bodapati et al. [24] | 0.9537 | MSENet |
| Abirami and Venkatesan [25] | 0.9552 | BCFA-based GAN |
| Satyanarayana et al. [26] | 0.9400 | CNN- based MCA |
| Deepak et al. [27] | 0.9540 | CNN (Deep feature fusion) + SVM |
| Khan et al. [28] | 0.9510 | Hybrid-NET |
| Agrawal et al. [29] | 0. 9640 | MultiFeNet |
| **Proposed Model** | **0. 9690** | **SKIPNet** |

**Discussion**

The performance of the proposed SKIPNet model is compared to various state-of-the-art brain tumor classification (BTC) methods, as outlined in the table. Prominently, Abirami and Venkatesan [25] developed a BCFA-based GAN model and attained 95.52% accuracy, which is comparable to the performance of the proposed model. Satyanarayana et al. [26] introduced a CNN-based MCA model that achieved an accuracy of 94.0%. While CNN architectures are known for their robust feature extraction capabilities, the relatively lower accuracy suggests that the model struggled to capture complex spatial patterns and lesion irregularities in brain MRI data. Similarly, Bodapati et al. [24] utilized MSENet, achieving an improved accuracy of 95.37%. Another approach, proposed by Deepak et al. [27], leveraged a combination of CNN architectures and SVMs, achieving an accuracy of 95.40%. This method relied on deep learning framework for effective capturing of texture details in brain MRI images. Pretrained DenseNet169 was coupled with modern machine learning algorithms for classification by Khan et al. [28] which classifies with an accuracy of 95.10%. Agrawal et al. [29] presented an advanced network by the name of MultiFeNet that achieved a higher accuracy of 96.40%. This demonstrates the potential of MultiFeNet to improve classification by enhancing spatial hierarchies, but the model still falls short compared to more advanced deep learning methods like SKIPNet. The proposed SKIPNet achieves a significant performance boost with an accuracy of 96.90%, outperforming all the discussed methods. This remarkable improvement can be attributed to SKIPNet's innovative integration of CNN Blocks and Spatial Attention Layers, which refine feature maps and emphasize critical spatial regions, ensuring robust and precise feature extraction. By leveraging deep learning techniques tailored for complex datasets, SKIPNet captures intricate spatial dependencies and lesion-specific details that other methods fail to address effectively. This performance underscores the model's efficacy and its potential to set a new benchmark in brain tumor classification.

## 5. Conclusion:

In conclusion, the proposed SKIPNet model demonstrates exceptional performance in brain tumor classification, achieving a superior accuracy of 96.90% compared to state-of-the-art methods. By integrating CNN Blocks with Spatial Attention Layers, SKIPNet effectively emphasizes critical spatial regions and refines feature extraction, enabling it to capture intricate spatial dependencies and lesion-specific details. Its ability to outperform traditional CNN-based approaches highlights its robustness, adaptability to complex datasets and demonstrates the strength of its end-to-end learning architecture. The model's innovative design not only ensures high classification accuracy but also maintains efficiency in terms of parameter count and memory requirements, making it well-suited for deployment in resource-constrained environments. SKIPNet establishes itself as a reliable and effective tool for brain tumor classification, paving the way for further advancements in medical image analysis.

**Declaration of competing interest**

The authors declare that they have no known competing financial interests or personal relationships that could have appeared to influence the work reported in this paper.

**Data availability**

The datasets used in this study are available publicly and the sources have been provided in the paper.